\documentclass[12pt]{iopart}	
\usepackage{graphicx}
\usepackage[usenames]{color}

\newcommand{\beq}{\begin{equation}}
\newcommand{\eeq}{\end{equation}}
\newcommand{\beqa}{\begin{eqnarray}}
\newcommand{\eeqa}{\end{eqnarray}}
\newcommand{\ba}{\begin{array}}
\newcommand{\ea}{\end{array}}

\begin{document}

\title{Mean-field 
 equations for cigar- and disk-shaped Bose and Fermi superfluids} 

\author{Camilo A. G.
Buitrago and S. K. Adhikari 
}
\address{Instituto de F\'{\i}sica Te\'orica, UNESP - S\~ao Paulo State
University,\\ 01.140-070 S\~ao Paulo, S\~ao Paulo, Brazil}

\begin{abstract} 

Starting from the three-dimensional (3D) time-dependent nonlinear
Gross-Pitaevskii equation for a Bose-Einstein condensate (BEC) and
density functional (DF) equation for a Fermi superfluid at the unitarity
and Bardeen-Cooper-Schrieffer (BCS)
 limits, we derive effective one- (1D) and two-dimensional (2D)
mean-field equations, respectively, for the dynamics of a trapped cigar-
and disk-shaped BEC and Fermi superfluid by using the adiabatic
approximation.  The reduced 1D and 2D equations for a cigar- and
disk-shaped Fermi superfluid have simple analytic nonlinear terms and at
unitarity produce results for stationary properties and non-stationary
breathing oscillation and free expansion in excellent agreement with the
solution of the full 3D equation.

\end{abstract}

\maketitle

\section{Introduction}

Many of the stationary and non-stationary properties of a trapped 
Bose-Einstein condensate (BEC) (or a Bose superfluid) 
can be satisfactorily explained  
by the 
mean-field Gross-Pitaevskii (GP) equation \cite{rev}.
However, there is no such mean-field equation in configuration space for 
the Fermi superfluid even 
after the experimental realization
of the crossover  from the 
Bardeen-Cooper-Schrieffer
(BCS) limit through unitarity to the Bose limit in a trapped Fermi 
superfluid \cite{cr,rmpfermi,K,Li}.

To study the  properties of stationary 
and 
non-stationary states of a trapped Fermi superfluid with equal number of 
paired spin-up and -down fermions, Adhikari and Salashnich developed a 
Galilei-invariant nonlinear density functional (DF)  equation 
valid from 
the weak-coupling 
BCS limit 
to unitarity \cite{DFAS,DFAS2}.  
The solution of the DF equation of  \cite{DFAS} 
yielded results for energy of a trapped Fermi superfluid in close 
agreement with those obtained from Monte Carlo calculations \cite{MC} 
not only in the BCS and unitarity limits but also along the 
BCS-unitarity crossover \cite{cr}.  


In actual experiments, an axially symmetric \cite{rev,rmpfermi,disk}, 
rather than a 
spherical,   trap is usually employed for the confinement of the 
BEC or 
Fermi superfluid.   In many situations of actual interest, the trap 
has an extreme geometry \cite{disk}, e.g., very strong  radial or 
axial  
confinement. Consequently, the superfluid is formed in the shape of a 
cigar or a disk and the solution of the nonlinear 
 equation in such cases 
deserves special attention. The cigar-shaped BEC or Fermi superfluid  is 
quasi 
one 
dimensional (1D) and the reduction of the full three-dimensional (3D)  
equation to accurate 1D form would be of advantage. Similarly, it would 
be beneficial to reduce the full 3D  equation to 2D form for the 
description of a disk-shaped BEC and Fermi superfluid. 

We propose a simple scheme for the 3D-1D and 3D-2D reduction of the GP 
equation for a BEC and the DF equation for a Fermi superfluid at the BCS 
and unitarity limits.
The 3D-1D reduction for a cigar-shaped BEC or a Fermi superfluid  is 
done in the
adiabatic approximation which assumes the essential dynamics to be
confined in the axial direction with the radial degrees of freedom
adjusting instantaneously to the minimum energy equilibrium
configuration compatible with the axial dynamics \cite{S1,M1}.  Under 
this
approximation the 3D wave function factorizes into an explicitly
time-dependent axial and time-independent radial parts which allows
for a formal reduction of the original time-dependent 3D nonlinear  
equation
into a time-dependent 1D and an auxiliary time-independent 2D equations.
The same is also true for a disk-shaped BEC or 
a Fermi 
superfluid
with the role of time-dependent 1D and auxiliary time-independent 2D
equations interchanged, e.g., the time-dependent equation is now 2D and
the auxiliary equation 1D in nature.

First we illustrate the present scheme through an application to a 
BEC described by the 
GP 
equation, where there are already several 
schemes \cite{S1,M1,S2,M2,O1,O2,O3,O4,O5,06} for 3D-1D and 3D-2D 
reductions.    Among  these, the reduction
schemes
of Salasnich {\it et al.} (SPR) \cite{S1,S2} and
Mu\~noz Mateo {\it et
al.} (MMD)
\cite{M1,M2} are the simplest to implement and
have been shown to be the most efficient \cite{M1}. 
We compare  the present scheme with  those of SPR 
\cite{S1,S2} and
MMD  
\cite{M1,M2} for a BEC. 
Recently, Adhikari and Salasnich (AS) \cite{njp} suggested one scheme
for such 3D-1D and 3D-2D reductions of the 3D DF equation for a Fermi
superfluid at the BCS and unitarity limits. 
{We apply our scheme to a Fermi superfluid and obtain 
distinct 3D-1D and 3D-2D reductions}.

In section \ref{II} we illustrate the 3D-1D and 3D-2D  reduction schemes in 
the 
case of the GP equation for a BEC. 
In section \ref{III} we consider the 
reduction 
schemes for the nonlinear DF equation of a Fermi superfluid at 
unitarity and BCS limits. In this case 
we reduce the 3D DF equation to 1D (for cigar-shaped superfluid) and 2D
(for disk-shaped superfluid) 
forms with analytic nonlinear terms. 
We  show that the reduced 1D and 2D 
equations for cigar- and 
disk-shaped Fermi superfluids are very effective for describing the
stationary states and  
non-stationary breathing oscillation and free expansion 
of the Fermi superfluid at unitarity in close 
agreement with the solution of the full 3D DF equation. 
Finally, in 
section 4 we give a summary and some concluding remarks. 

\section{3D-1D and 3D-2D Reductions of the GP equation}

\label{II}

We first apply our approach to a 
BEC described by the  
GP equation.
The GP equation for $N$ bosons of mass $m$ is written as 
\cite{rev}
\begin{equation}\label{e1}
\left[-i\hbar\frac{\partial }{\partial \hat t}
-\frac{\hbar^2\nabla_{\bf \hat r}^2}{2m}
+V({\bf \hat r})+  \frac{4\pi \hbar^2 \hat a N}{m}|\Psi({\bf r})|^{2}
\right]\Psi({\bf \hat   r},t)=0,
\end{equation}
with  $\hat a$ the atomic scattering length, 
normalization $\int |\Psi|^2 d^3{\bf \hat r}=1$,  
$V({\bf \hat r})= m\omega^2(\hat \rho^2+\lambda^2 
\hat z^2)/2$ 
the harmonic trapping potential with frequencies $\omega$ and $\lambda 
\omega$ in radial ($\hat \rho$) and axial ($\hat z$) directions $({\bf 
\hat r}\equiv 
\hat \rho,\hat z).$  Employing 
dimensionless harmonic oscillator units $t=\hat t \omega, 
{\bf r= \hat r}/a_\rho, 
z=\hat z/a_\rho, \rho =\hat \rho/a_\rho, a=\hat 
a/a_\rho, 
\psi  =\Psi 
a_\rho^{3/2}, a_\rho=\sqrt{ \hbar/m\omega}$,    (\ref{e1}) can be 
written as
\begin{equation}\label{e2}
\left[-i\frac{\partial }{\partial  t}
-\frac{\nabla_{\bf r}^2}{2}+\frac{\rho^2+\lambda^2 z^2}{2}
+4\pi a N|\psi|^{2}
\right]\psi({\bf \bf r},t)=0,
\end{equation}
with normalization $\int |\psi|^2 d^3{\bf  
r}=1.$

In the limits of very small (cigar-shaped trap) and very large 
(disk-shaped trap) $\lambda$,  the length scales in the axial 
and radial
directions are very different. Consequently,  the correlations between 
these two
directions can be neglected and the condensate wave function 
could be factorized in variables $\rho$ and $z$ \cite{S1,M1}. In 
the 
case of cigar-shaped traps  the dynamics takes place in the axial 
direction. The opposite happens in case of a disk-shaped 
trap.   
For a stationary solution $ \psi({\bf r})$ of  (\ref{e2}) in a 
cigar-shaped trap one can define a linear 
density $\phi^2(z)\equiv \int d^2 \rho\psi^2({\bf r})$. Similarly, 
for a disk-shaped trap  one can define a radial density 
$\varphi^2(\rho)\equiv \int d z\psi^2({\bf r})$. 

\subsection{3D-1D reduction for a cigar-shaped BEC}

For a cigar-shaped trap, the above consideration leads to the 
factorization \cite{M1}
\begin{equation}\label{e3}
\psi({\bf r},t)=\varphi(\rho,n_1(z,t))\phi(z,t),
\end{equation}
where linear density $n_1$ is defined as $n_1(z,t)\equiv$ $ 
N|\phi(z,t)|^2 = N\int d^2 
\rho 
|\psi|^2$ and normalizations $\int d^2\rho |\varphi(\rho,n_1)|^2= \int 
dz 
|\phi(z,t)|^2=1$. We {\it assume} 
that the function $\varphi(\rho,n_1(z,t))$ has no explicit $t$ or $z$ 
dependence and hence these derivatives do not act on this function.
This is a good approximation, in general, as we find in numerical studies.  
The substitution of  (\ref{e3}) in  (\ref{e2}) then 
leads to
\begin{eqnarray}\label{e4}
\fl \varphi(\rho,n_1)\left[i\frac{\partial }{\partial t}+\frac{1}{2}
\frac {\partial^2 }{\partial z^2}-\frac{1}{2}\lambda^2z^2
\right]\phi(z,t)
{
=\phi(z,t)\left[-\frac{1}{2}\nabla_\rho^2+\frac{1}{2}
{\rho^2}+ 
4\pi a  n_1|\varphi|^{2}\right]\varphi(\rho,n_1).}
\end{eqnarray}
Multiplying  (\ref{e4}) by $\varphi^*(\rho,n_1)$ and integrating in 
$\rho$ this equation can be rewritten as  \cite{M1}
\begin{eqnarray}\label{e5}
&&\left[i\frac{\partial }{\partial t}+\frac{1}{2}
 \frac{\partial^2 }{\partial z^2}-\frac{1}{2}
\lambda^2z^2-\mu_\rho(n_1)
\right]\phi(z,t)=0,\\ 
&&{\mu_\rho(n_1)=\int d^2\rho\varphi^*\left[  
-\frac{1}{2}\nabla_\rho^2+\frac{1}{2}{\rho^2}+
4\pi a n_1 |\varphi|^2\right]\varphi,}
\label{e7}  
\end{eqnarray}
where  $\mu_\rho(n_1)$ is the chemical 
potential emerging from the 2D GP equation:
\begin{eqnarray}
{\left[-\frac{1}{2}\nabla_\rho^2+\frac{1}{2}{\rho^2}+
4\pi a n_1  
|\varphi|^{2}- \mu_\rho(n_1)
\right]\varphi(\rho,n_1)=0.} 
 \label{e6}
\end{eqnarray}
We have decoupled the essential axial 
($z$) 
and nonessential radial ($\rho$)
degrees of freedom.  
The solution of the time-independent radial GP equation (\ref{e6}) 
leads to the chemical potential $\mu_\rho(n_1)$ given by  
(\ref{e7}), which is the nonlinear term of the axial GP equation 
(\ref{e5}).

The form of the chemical potential $\mu_\rho(n_1)$ of  
(\ref{e6}) is known in the small and large $N$ limits. In the 
small $N$ weak-coupling limit, the wave function can be  approximated by 
the following normalized  
Gaussian form \cite{PG}
\begin{equation}
\varphi(\rho,n_1)=\exp[-\rho^2/(2\alpha^2)]
/(\sqrt{\alpha^2 
\pi}), \label{g2D}
\end{equation}
where $\alpha$ is the width. 
With this wave form the chemical 
potential of  (\ref{e7}) 
becomes 
\begin{eqnarray}\label{e10}
{\mu_\rho(n_1)= 
\biggr(\frac{\alpha^2}{2}+\frac{1}{2\alpha^2}\biggr)+\frac{ 2a 
n_1}{\alpha^2}
.}
\end{eqnarray}
In the large $N$ Thomas-Fermi (TF) limit, as 
$an_1 \to  \infty ,$ the kinetic 
energy 
gradient 
operator 
in  (\ref{e6}) can be neglected and this equation has analytic 
solution. The normalization condition of the TF wave function 
leads to  \cite{S1,M1}
\begin{eqnarray}\label{e11}
{\mu_\rho(n_1)=\sqrt{4an_1}}.
\end{eqnarray}

We suggest  the following simple interpolation formula for 
$\mu_\rho(n_1)$ valid for small to large  $an_1$ 
incorporating the limiting values (\ref{e10})
and (\ref{e11})
\begin{eqnarray}\label{e12}
{\mu_\rho(n_1)=\frac{1}{2\alpha^2}-
\frac{\alpha^2}{2}+\sqrt{\alpha^4+4an_1}}
\end{eqnarray}
to be used in  (\ref{e5}), 
where $\alpha$ is taken as  a fixed constant for all $n_1$.
In the weak-coupling $an_1 \to 0$ limit,  (\ref{e6}) reduces to the 
Schr\"odinger equation for a 2D harmonic oscillator with the exact 
solution (\ref{g2D}) with $\alpha = 1$. 
Then  (\ref{e12}) is a good 
approximation to  (\ref{e7}) for $\alpha=1$. 
For slightly larger 
values of $an_1$  (\ref{e12}) continues to be a good approximation to 
   (\ref{e7}), but with a slightly smaller value of 
$\alpha$. Motivated by this, we take a slightly smaller value of 
$\alpha$ in  (\ref{e12}). For  large  $an_1$, 
 (\ref{e12}) has the correct TF limit independent of the value of 
$\alpha$ employed.     
This
flexibility  in introducing a width $\alpha\approx 1$ 
(slightly different from $\alpha = 1$) 
in   (\ref{e12}) will  be fundamental in making the 1D model 
equation (\ref{e5}) 
a faithful approximation to the 3D GP equation (\ref{e2}) for a 
cigar-shaped condensate for 
all  $an_1$.

By construction, approximation (\ref{e12}) satisfies the weak-coupling 
and 
TF limits (\ref{e10}) and (\ref{e11}), respectively, for small and large 
$an_1$. 
The approximation of  MMD is \cite{M1} 
\begin{eqnarray}\label{e8}
\mu_\rho(n_1)=\sqrt{1+4an_1},
\end{eqnarray}
whereas SPR suggested \cite{S1}
\begin{eqnarray}\label{e9}
\mu_\rho(n_1)=\frac{1+3an_1}{\sqrt{1+2an_1}}.
\end{eqnarray}

Here we calculate the chemical potential 
$\mu_\rho(n_1)$ of the three approaches and compare  with the 
precise results for $\mu_\rho(n_1)$ obtained from a 
numerical solution of  (\ref{e6}). 
(All numerical results presented in this paper are obtained with the 
imaginary-time propagation scheme after a discretization by the 
Crank-Nicholson method using the FORTRAN programs provided in \cite{bo}, 
the details of which are described there. 
The numerical simulations for the dynamical breathing 
oscillation and free expansion for a Fermi superfluid at unitarity 
reported 
in Secs. IIIC and IIID, respectively,  were performed with the real-time 
propagation scheme after a 
discretization by the Crank-Nicholson method.)
Our finding is exhibited in table 
\ref{btable1} for different  $an_1$ and
for $\alpha=0.985$ together with 
those obtained from the MMD \cite{M1} and SPR \cite{S1} schemes.

\begin{table}[!ht]
\begin{center}
\caption{Chemical potential $\mu_\rho(n_1)$ of  (\ref{e7}) for
different
$an_1$ obtained from an accurate numerical solution of  (\ref{e6}),
and from  (\ref{e12})    (Present, $\alpha=0.985$), (\ref{e8})
(MMD) \cite{M1},  and
(\ref{e9})
(SPR) \cite{S1}.
}
\label{btable1}
\begin{tabular}{|r|r|r|r|r|}
\hline
$an_1$ & Numerical  & MMD   &
 SPR    &   Present   \\
\hline
  0  &  1  & 1  & 1  &  1.00045
\\
0.2  &   1.346427      & 1.34164  &  1.35225 &  1.34983   \\
  1  &  2.257135  & 2.23607  & 2.30940 & 2.25314 \\
 10  &    6.432456  &  6.40312  &6.76475   & 6.42877 \\
  100  & 20.04320  & 20.0250  &  21.2309 & 20.0538\\
\hline
\end{tabular}
\end{center}
\end{table}


\subsection{3D-2D reduction for a disk-shaped BEC}

For a disk-shaped trap the adiabatic approximation  leads to 
the 
factorization \cite{S1,M1}
\begin{equation}\label{e13}
\psi({\bf r},t)=\varphi(\rho,t)\phi(z,n_2(\rho,t)),
\end{equation}
where  the surface density $n_2$ is defined as $n_2(\rho,t)\equiv 
N|\varphi(\rho,t)|^2 = N\int dz 
|\psi|^2$ and normalizations $\int d^2\rho |\varphi(\rho,t)|^2= \int dz 
|\phi(z,n_2)|^2=1$. The substitution of  (\ref{e13}) in  
(\ref{e2})
leads to
\begin{eqnarray}\label{e14}
\fl \phi(z,n_2)\left[i\frac{\partial }{\partial 
t}+\frac{1}{2}\nabla_\rho^2
-\frac{1}{2}\rho ^2
\right]\varphi(\rho,t)
=\varphi(\rho,t)\left[-\frac{1}{2}\frac{\partial^2}{\partial 
z^2}+\frac{1}{2}\lambda^2{z^2}+ 4\pi a n_2
|\phi|^2\right]\phi(z,n_2).
\end{eqnarray}
Multiplying  (\ref{e14}) by $\phi^*(z,n_2)$ and integrating in 
$z$, this equation can be rewritten as the set of equations 
\begin{eqnarray}\label{e15}
\left[i\frac{\partial }{\partial t}+
\frac{1}{2}\nabla_\rho^2
-\frac{1}{2}\rho^2-\mu_z(n_2)
\right]\varphi(\rho,t)=0,\\
 \label{e16}
\left[-\frac{1}{2}\frac {\partial^2 }{\partial z^2}
+\frac{\lambda ^2z^2}{2}+4\pi a
 n_2|\phi|^{2}-\mu_z(n_2)
\right]\phi(z,n_2)=0,\\
\mu_z(n_2)=\int dz\phi^*\left[  -\frac{1}{2}\frac {\partial^2 }{\partial 
z^2}
+\frac{\lambda^2z^2}{2}+
4\pi a  n_2|\phi|^{2}\right]\phi.\label{e17}
\end{eqnarray}
It is convenient to introduce scaled variables $\bar z=z/a_z, \bar\phi 
={\sqrt a_z}\phi,$ and $\bar \mu_z=\mu_z a_z^2$ with 
$a_z=\sqrt{1/\lambda}$, when  (\ref{e16}) and (\ref{e17}) become 
\begin{eqnarray}\label{e18}
&&\left[-\frac{1}{2}\frac {\partial^2 }{\partial \bar z^2}
+\frac{\bar z^2}{2}+4\pi a 
 a_z n_2|\bar \phi|^{2}-\bar \mu_z(n_2)
\right]\bar \phi(z,n_2)=0,  \\&&
\bar \mu_z(n_2)=\int d\bar z\bar \phi^*\left[  -\frac{1}{2}\frac 
{\partial^2 
}{\partial \bar z^2}
+\frac{\bar z^2}{2}+
4\pi a a_z n_2|\bar \phi|^{2}\right]\bar \phi. \label{e19}
\end{eqnarray}

The form of the chemical potential $\bar \mu_z(n_2)$ of  
(\ref{e18}) is known in the small and large $N$ limits. In 
the 
small $N$ weak-coupling limit the wave function can be  approximated by 
the following normalized 
Gaussian form \cite{PG}
\begin{equation}
\bar \phi(\bar z,n_2)=\exp[-\bar z^2/(2\eta^2)]/(\eta^2 
\pi)^{1/4}, \label{g1D}
\end{equation}
where $\eta$ is the width. 
With this wave form the chemical 
potential of  
(\ref{e19}) 
becomes 
\begin{equation}\label{e20}
\bar \mu_z(n_2)= 
\biggr(\frac{\eta^2}{4}+\frac{1}{4\eta^2}\biggr)+
2aa_zn_2\frac{\sqrt{2\pi}}{\eta}.
\end{equation}
In the large $N$ limit 
the normalization condition of the TF wave function 
leads to \cite{M1}
\begin{equation}\label{e21}
\bar \mu_z(n_2)=(3\pi a a_z n_2/\sqrt 2 )^{2/3}.
\end{equation}

For a disk-shaped BEC we suggest  the following simple interpolation 
formula for
$\bar \mu_z(n_2)$ incorporating the limiting values (\ref{e20})
and (\ref{e21})
\begin{equation}\label{e22}
\bar \mu_z(n_2)= \frac{1}{4\eta^2}-
\frac{(\pi-1) \eta^2}{4}+
\left[
\left(
\frac{\pi\eta^2}{4}
\right)^{3/2}
+\frac{3\pi a a_zn_2 }{\sqrt 2}\right]^{2/3},
\end{equation}
valid for all $aa_zn_2$,
where $\eta$ is taken as a fixed constant for all 
$aa_zn_2$.
The flexibility in introducing a width $\eta$ slightly different from 
1 in  (\ref{e22}) will make the 2D model equation 
(\ref{e15}) an accurate approximation to the 3D GP equation (\ref{e2}) 
for 
a disk-shaped condensate for all $aa_z^2n_2$. 

Instead of taking $\eta$ as a constant, SPR \cite{S1} solve 
(\ref{e18}) variationally \cite{S1,PG}
with the Gaussian ansatz
(\ref{g1D}) and determine the width parameter $\eta$ via
\begin{equation}\label{e23}
\eta^4-2\sqrt{2\pi}\eta aa_zn_2-1=0.
\end{equation}
The solution of the nonlinear
 (\ref{e23}) when substituted into  (\ref{e20})
yields the desired $\bar\mu_z$ \cite{S1} through a procedure far
complicated than
the analytic formulae (\ref{e22}).

The approximation scheme of MMD is quite involved but does not
require the solution of a nonlinear variational equation like SPR. They
calculate $\bar \mu_z(n_2)$ via \cite{M1}
\begin{equation}\label{e26}
\bar \mu_z(n_2)=\frac{1}{8}[(\eta+\sqrt{\eta^2-\zeta^6 })^{1/3}
+(\eta-\sqrt{\eta^2-\zeta^6 })^{1/3}-\zeta]^2
\end{equation}
where $\eta=4+6\zeta-\zeta^3+24\pi x$ and $\zeta=(\kappa-1)$
with
\begin{equation}\label{theta}
\kappa^{-1}=\sqrt{2/\pi}+\Theta(x-0.1)(1-\sqrt{2/\pi})[1-(10x)^{-1/5}]
\end{equation}
where $x\equiv aa_z  n_2$, and the Heaviside step function 
$\Theta(x-0.1)=0,$ for $x<0.1$, and 
$=1$ for $x>0.1$. It is realized that the function (\ref{theta}) is not 
analytic in $x$.

\begin{table}[!ht]
\begin{center}
\caption{Chemical potential $\bar \mu_z(n_2)$ of
 (\ref{e19}) for different
$aa_zn_2$ obtained from an accurate numerical solution of 
(\ref{e18}),
and from  (\ref{e22})    (Present, $\eta=0.95$), (\ref{e26})  
(MMD) \cite{M1},
and
(\ref{e23})
(SPR) \cite{S1}.
}
\label{btable2}
\begin{tabular}{|r|r|r|r|r|}
\hline
$aa_zn_2$ & Numerical  & MMD   &
 SPR    &   Present   \\
\hline
  0  &  0.5  & 0.5  & 0.5  &    0.50263
\\
0.2  &    1.348783     & 1.31186  &  1.36149 &   1.34376   \\
  1  &   3.599892   & 3.50165 & 3.69266   &   3.54355 \\
  10  &    16.45405    & 16.2547     & 17.0012  & 16.3294 \\
  100  &  76.30080      & 75.9963  &78.8855   & 76.1358\\
\hline
\end{tabular}
\end{center}
\end{table}

Now we calculate the chemical potential 
$\bar \mu_z(n_2) $ of the three approaches and compare them with the 
precise result for $\bar \mu_z(n_2) $  from a numerical solution 
of  (\ref{e18}). 
The results are shown in table \ref{btable2} for 
different  $aa_zn_2$ along with those from the MMD and SPR schemes. 
 After a small experimentation 
the constant  $\eta$ in  (\ref{e22}) was fixed at $\eta=0.95$ for 
all $aa_zn_2$. 

\section{3D-1D and 3D-2D reductions of Fermi superfluid DF equations}

\label{III}

We consider a Gallilei-invariant density-functional (DF) formulation for 
a 
trapped Fermi superfluid at BCS and unitarity limits described  by
\cite{DFAS,DFAS2}
\begin{equation}\label{xe1}
\left[-i\hbar\frac{\partial }{\partial \hat t}
-\frac{\hbar^2\nabla_{\bf \hat r}^2}{4m}
+2V({\bf \hat r})+  2^{2/3}\chi \frac{2\hbar^2}{m}|\Psi({\bf 
r})|^{4/3}
\right]\Psi({\bf \hat   r},t)=0,
\end{equation}
with 
$\chi=(3\pi^2)^{2/3}\xi/2$ ($\xi=1$ at the BCS limit and $\xi 
=0.44$ at unitarity \cite{MCCOL}), 
$N$ the number of fermions, and 
normalization $\int |\Psi|^2 d^3{\bf \hat r}=N/2$ ($|\Psi|^2$ is the 
density of Fermi pairs), 
$m$ the mass of an atom, 
$V({\bf \hat r})= m\omega^2(\hat \rho^2+\lambda^2 
\hat z^2)/2$ 
the harmonic trapping potential with frequencies $\omega$ and $\lambda 
\omega$ in radial ($\hat \rho$) and axial ($\hat z$) directions $({\bf 
\hat r}\equiv 
\hat \rho,\hat z).$  The fully-paired 
Fermi superfluid is assumed to be composed of 
spin-half fermions with 
an equal number of spin-up and -down components. 
Employing 
dimensionless units $t=\hat t \omega, 
{\bf r= \hat r}/a_\rho, 
z=\hat z/a_\rho, \rho =\hat \rho/a_\rho, a=\hat a/a_\rho, 
\psi \sqrt{N/2} =\Psi 
a_\rho^{3/2}, a_\rho=\sqrt{\hbar/m\omega}$,    (\ref{xe1}) can be 
written as
\begin{equation}\label{xe2}
\left[-i\frac{\partial }{\partial  t}
-\frac{\nabla_{\bf r}^2}{4}+{\rho^2+\lambda^2 z^2}
+2\chi N^{2/3}|\psi|^{4/3}
\right]\psi({\bf \bf r},t)=0,
\end{equation}
with $\int |\psi|^2 d^3{\bf  r}=1 (|\psi|^2$ is the density of Fermi 
atoms). 


\subsection{3D-1D reduction for a cigar-shaped Fermi superfluid}

For  a cigar-shaped trap, we consider  the 
factorization (\ref{e3}).
We substitute   (\ref{e3}) in  (\ref{xe2}) and 
multiply the resultant equation  by $\varphi^*(\rho,n_1)$ and 
integrate in 
$\rho$ to get 
\begin{eqnarray}\label{xe5}
&&\left[i\frac{\partial }{\partial t}+\frac{1}{4}
 \frac{\partial^2 }{\partial z^2}-\lambda^2z^2-\mu_\rho(n_1)
\right]\phi(z,t)=0,
\label{xe7}  
\end{eqnarray}
where  $\mu_\rho(n_1)$ is  the chemical 
potential emerging from the following 2D DF equation
\begin{eqnarray}
\left[-\frac{\nabla_\rho^2}{4}+{\rho^2}+
2\chi 
 n_1^{2/3}|\varphi|^{4/3}- \mu_\rho(n_1)
\right]\varphi(\rho,n_1)=0. 
 \label{xe6}
\end{eqnarray}

In the 
small $N$ weak-coupling limit, the wave function $\varphi(\rho,n_1)$
can be  approximated by 
the  normalized  
Gaussian form \cite{PG}  (\ref{g2D}).
With this  wave form 
the chemical 
potential of  (\ref{xe6}) 
becomes 
\begin{eqnarray}\label{xe10}
\mu_\rho(n_1)= 
\biggr({\alpha^2}+\frac{1}{4\alpha^2}\biggr)+
\frac{6\chi}{5}
\frac{n_1^{2/3}}{\alpha^{4/3}\pi^{2/3}}.
\end{eqnarray}
 Chemical potential  $\mu_\rho(n_1)$ 
of  (\ref{xe10}) is 
consistent with  (4.7) of AS \cite{njp}. 
In the large $N$ TF limit  
the normalization condition of the TF wave function 
leads to \cite{S1,M1}
\begin{eqnarray}\label{xe11}
\mu_\rho(n_1)=\left[
\left(\frac{5n_1}{2\pi}\right)^{2/3}
2\chi 
\right]^{3/5}\approx 1.38336 n_1^{2/5}\chi^{3/5}.
\end{eqnarray}
Chemical potential (\ref{xe11}) is approximately equal to the 
chemical potential  in the  
large $N$ 
limit of the corresponding model (4.10) of AS \cite{njp} which 
yields in present notation 
\begin{equation}\label{xe11x}
\mu_\rho(n_1)=\frac{7}{5}\frac{(6\chi)^{3/5}n_1^{2/5}}
{(5\pi^2)^{1/5}2^{2/5}}
\approx 1.42545 n_1^{2/5}\chi^{3/5}.
\end{equation}
Here we use the following simple interpolation formula for 
$\mu_\rho(n_1)$ incorporating the limiting values (\ref{xe10})
and (\ref{xe11})
\begin{eqnarray}\label{xe12}
\mu_\rho(n_1)=
\frac{1}{4\alpha^2}-\frac{3\alpha^2}{2} 
 +
\left[\left( \frac{5\alpha^2}{2}   \right)^{5/3}
+2\chi \left(\frac{5n_1}{2\pi}\right)^{2/3}
\right]^{3/5}.
\end{eqnarray}
where $\alpha$ is taken as  a fixed constant for all $n_1$. 


AS \cite{njp} solve Eq. (\ref{xe6}) variationally and 
obtain for the width $\alpha$
\begin{equation}\label{xx10}
\alpha^4= 
\frac{1}{4}+\frac{12\chi}{25}\left(\frac{n_1\alpha}{\pi}\right)^{2/3}.
\end{equation}

\begin{table}[!ht]
\begin{center}
\caption{Chemical potential 
$\mu_\rho(n_1)$ of  (\ref{xe7}) for 
different 
$n_1$ for a Fermi superfluid at unitarity  
($\xi=0.44$)
obtained from an accurate numerical solution of  (\ref{xe6}), 
from  (\ref{xe10})  and (\ref{xx10}) (AS)
\cite{DFAS},
and from  (\ref{xe12})    (Present, $\alpha=1/\sqrt2$, and $ 
0.98/\sqrt 2$). 
}
\label{table1}
\begin{tabular}{|r|r|r|r|r|}
\hline
 $n_1$ & Numerical    &
 AS    &   Present  &   Present  \\
    &    &   & $\alpha=1/\sqrt 2$   &  $\alpha=0.98/\sqrt 2$  \\
\hline
  0  &  1  & 1  & 1  &  1.00081 \\
0.1  & 1.37401          &     1.37619 & 1.36784  &1.37667\\
1  &    2.43893       &    2.46380  & 2.39751  &  2.41810\\
 10 & 5.56376           &   5.69170 &  5.45891 & 5.49147 \\
100  &13.70413           &    14.09704  & 13.5463  & 13.5867  \\
1000  &34.29934           &   35.33109  & 34.1057  & 34.1506\\
\hline
\end{tabular}
\end{center}
\end{table}

\begin{figure}

\begin{center}
\includegraphics[width=.49\linewidth]{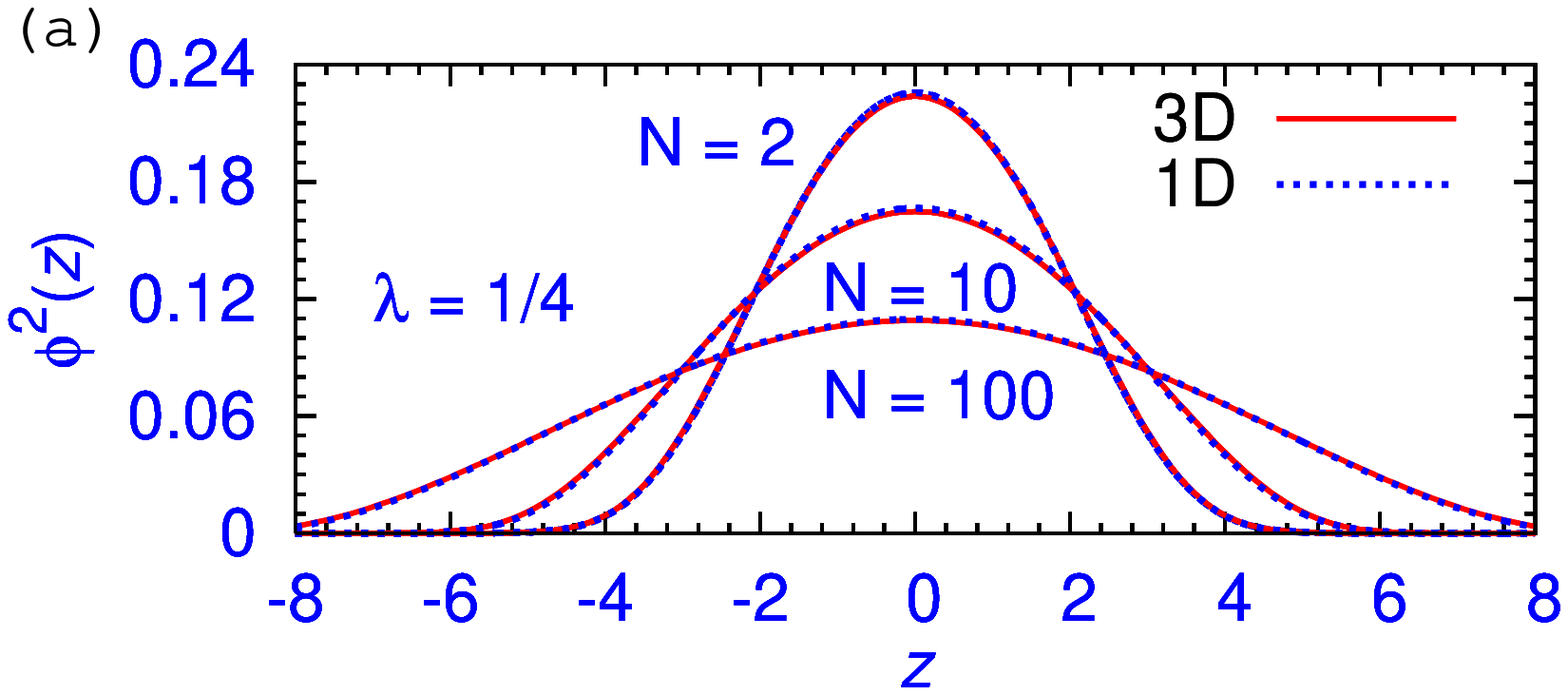}
\includegraphics[width=.49\linewidth]{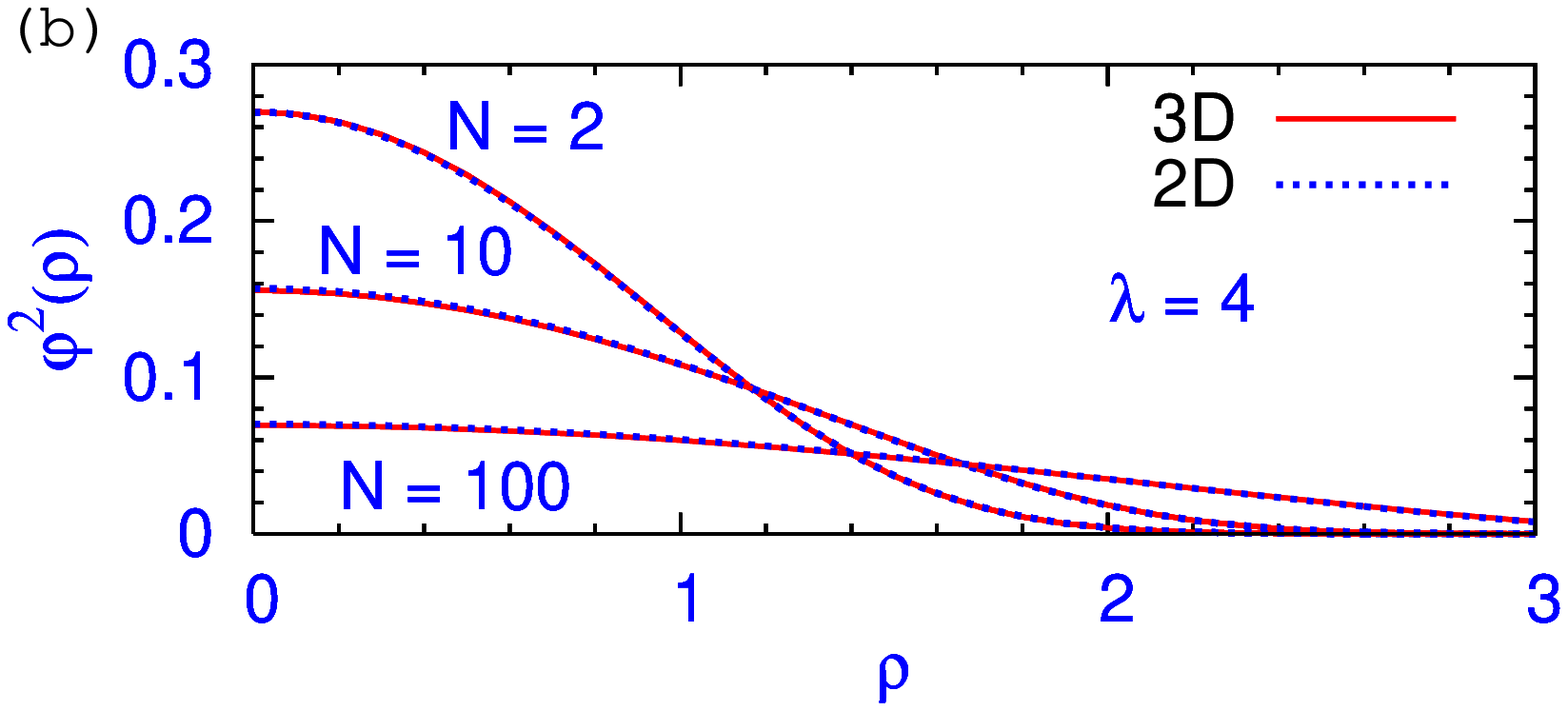}
\end{center}

\caption{(Color online) (a) The linear density $\phi^2(z)$ of a Fermi 
superfluid at unitarity (with $\xi =0.44$) vs. $z$
(both in dimensionless units)
calculated from 
the 3D DF  equation (\ref{xe2}) and the 1D model  (\ref{xe5}) and 
(\ref{xe12}) for 
$\lambda =1/4, \alpha =0.98/\sqrt 2$ and  $N=2$,  10, and  100.  
(b) The radial density $\varphi^2(\rho)$ of a Fermi
superfluid at unitarity (with $\xi =0.44$) vs. $\rho$
(both in dimensionless units)
calculated 
from 
the 3D DF  equation (\ref{xe2}) and the 2D model  (\ref{xe15}) 
and 
(\ref{xe22}) for 
$\lambda =4, \eta =0.92/\sqrt 2$ and  $N=2$,  10, and  100.  
}\label{fig2}
\end{figure}

Here we calculate the chemical potential 
$\mu_\rho(n_1)$ obtained from  (\ref{xe12}) for different 
$\alpha$ 
and from  (4.7) and (4.8) of  \cite{njp}
 and compare  with the 
precise result for $\mu_\rho(n_1)$  from a 
numerical solution of  (\ref{xe6}). 
Our finding is exhibited in table 
\ref{table1} for different  $n_1$ together with those from the AS 
\cite{njp} scheme. 
After a small 
experimentation it was found that the best overall result from 
 (\ref{xe12}) was found for 
$\alpha=0.98/\sqrt 2$.  

Now to see how well the effective 1D equations (\ref{xe5})
and (\ref{xe12}) reproduce the linear density $\phi^2(z)$ of a 
cigar-shaped Fermi superfluid we plot in figure  \ref{fig2} (a) the 1D density 
calculated from  (\ref{xe5})
and (\ref{xe12}) and the same calculated from the full 3D  
(\ref{xe2}) for $\lambda =1/4$ and $N=2,10,100.$ The excellent  
agreement 
between the two sets of results for $\lambda $ as large as 1/4
demonstrates the usefulness of the 
present 1D model equations.

\subsection{3D-2D reduction for a disk-shaped Fermi superfluid}

In the case of a disk-shaped trap we consider  the 
factorization (\ref{e13}) and 
substitute  it  in  
(\ref{xe2}),
multiply the resultant equation  by $\phi^*(z,n_2)$ and integrate 
in 
$z$ to get 
\begin{equation}\label{xe15}
\left[i\frac{\partial }{\partial t}+\frac{1}{4}\nabla_\rho^2
-\rho^2-\mu_z(n_2)
\right]\varphi(\rho,t)=0,
\end{equation}
where $\mu_z(n_2)$ is the chemical potential emerging from the following 
1D DF equation
\begin{equation}
 \label{xe16}
\left[-\frac{1}{4}\frac {\partial^2 }{\partial z^2}
+{\lambda ^2z^2}+2\chi
 n_2^{2/3}|\phi|^{4/3}-\mu_z(n_2)
\right]\phi(z,n_2)=0.
\end{equation}
It is convenient to introduce scaled variables $\bar z=z/a_z, \bar\phi 
={\sqrt a_z}\phi,$ and $\bar \mu_z=\mu_z a_z^2$ with 
$a_z=\sqrt{1/\lambda}$, when  (\ref{xe16}) 
becomes 
\begin{eqnarray}\label{xe18}
&&\left[-\frac{1}{4}\frac {\partial^2 }{\partial \bar z^2}
+{\bar z^2}+2\chi
 a_z^{4/3} n_2^{2/3}|\bar \phi|^{4/3}-\bar \mu_z(n_2)
\right]\bar \phi(z,n_2)=0,  \\&&
\bar \mu_z(n_2)=\int d\bar z\bar \phi^*\left[  -\frac{1}{4}\frac 
{\partial^2 
}{\partial \bar z^2}
+{\bar z^2}+
2\chi a_z^{4/3} n_2^{2/3}|\bar \phi|^{4/3}\right]\bar \phi. \label{xe19}
\end{eqnarray}

The form of the chemical potential $\bar \mu_z(n_2)$ of 
(\ref{xe18}) is known in the small and large $N$ limits. In 
the 
small $N$ weak-coupling limit the wave function $\bar \phi(\bar z,n_2)$
can be  approximated by 
the  normalized 
Gaussian form \cite{PG} (\ref{g1D}).
With this wave form the chemical 
potential of  
(\ref{xe19}) 
becomes 
\begin{equation}\label{xe20}
\bar \mu_z(n_2)= 
\biggr(\frac{\eta^2}{2}+\frac{1}{8\eta^2}\biggr)+
\frac{2\chi (n_2a_z^2)^{2/3}}{\eta^{2/3}
\pi^{1/3}}\sqrt{3\over 5}.
\end{equation}
Chemical potential  
$\bar \mu_z(n_2)$ of  
(\ref{xe20}) is consistent with 
 (5.7) of AS \cite{njp}.  
In the large $N$ TF limit 
the normalization condition of the TF wave function 
leads to \cite{M1}
\begin{equation}\label{xe21}
\bar \mu_z(n_2)= \left[\frac{8\chi    n_2^{2/3}a_z^{4/3}
}{(3\pi)^{2/3}} 
\right]^{3/4}\approx 1.54947 \chi^{3/4}a_z\sqrt{n_2}.
\end{equation}
Chemical potential (\ref{xe21}) is approximately equal to the
chemical potential  in the
large $N$
limit of the corresponding model (5.10)  of AS \cite{njp} which
yields in present notation
\begin{equation}\label{xe21x}
\bar \mu_z(n_2)=\frac{12a_z 3^{3/8}\chi^{3/4}\sqrt n_2
}{5\sqrt 2  \pi^{1/4}5^{1/8}}
\approx 1.57383 \chi^{3/4}a_z\sqrt n_2.
\end{equation}
Here we use the following  interpolation formula for 
$\bar \mu_z(n_2)$ incorporating the limiting values (\ref{xe20})
and (\ref{xe21})
\begin{eqnarray}\label{xe22}
\bar \mu_z(n_2)= \left(\frac{\eta^2}{2}+\frac{1}{8\eta^2} 
\right)-\frac{15^{3/2}\eta^2}{9\pi}
+\left[ 
\left(   \frac{15^{3/2}\eta^2  }{9\pi}  \right)^{4/3}+
8\chi   \frac{(n_2a_z^2)^{2/3}
}{(3\pi)^{2/3}}
\right]^{3/4}.
\end{eqnarray}
where $\eta$ is taken as a fixed constant for all $a_z^2n_2$. 

AS \cite{njp} solve Eq. (\ref{xe18}) variationally and obtain for the 
width $\eta$ 
\begin{equation}\label{xy22}
\eta^4= 
\frac{1}{4}+\frac{4\chi}{5}\frac{(a_z^2n_2\eta^2)^{2/3}}{\pi^{1/3}}\sqrt{3\over 
5}.
\end{equation}



\begin{table}[!ht]
\begin{center}
\caption{Chemical potential $\bar \mu_z(n_2)$ of 
 (\ref{xe19}) for different
$a_z^2n_2$ for a Fermi superfluid at unitarity
($\xi=0.44$)
obtained from an accurate numerical solution of  
(\ref{xe18}), from  (\ref{xe20})  and (\ref{xy22}) (AS)
\cite{DFAS}, 
and from  (\ref{xe22})    (Present, $\eta=1/\sqrt 2$ and 
$\eta=0.92/\sqrt 2$).
}
\label{table2}
\begin{tabular}{|r|r|r|r|r|}
\hline
$a_z^2n_2$ & Numerical    &
 AS    &   Present  &  Present \\
    &    &  & $\eta=1/\sqrt 2$ &$\eta=0.92/\sqrt 2$  \\
\hline
  0  &  0.5  & 0.5  & 0.5  &  0.50696\\
0.1  & 1.071781         &   1.07375 & 1.05892&  1.08883  \\
1  & 2.79861         & 2.82290  &  2.72155& 2.78109  \\
 10 & 8.59782         & 8.72108  & 8.41456&  8.50454  \\
100    & 27.09534         &  27.51645  &26.8149 & 26.9282  \\
500  & 60.56659         &  61.51645  &  60.2317   &60.3571 \\
\hline
\end{tabular}
\end{center}
\end{table}

As in the 3D-1D reduction, now we calculate the chemical potential 
$\bar \mu_z(n_2) $ of the three approaches and compare them with the 
precise result for $\bar \mu_z(n_2) $ obtained from a numerical solution 
of  (\ref{xe18}). The results are shown in table \ref{table2} for 
different  $a_z^2n_2$ along with those from the AS scheme \cite{njp}. 

Now to see how well the effective 2D equations (\ref{xe15})
and (\ref{xe22}) reproduce the radial density $\varphi^2(\rho)$ of a 
disk-shaped Fermi superfluid we plot in figure  \ref{fig2} (b) the 2D density 
calculated from  (\ref{xe15})
and (\ref{xe22}) and the same calculated from the full 3D  
(\ref{xe2}) for $\lambda =4$ and $N=2,10,100.$ The excellent agreement 
between the two sets of results for $\lambda$ as small as 4 
demonstrates the usefulness of the 
present 2D model equations. 

\begin{figure}

\begin{center}
\includegraphics[width=.49\linewidth]{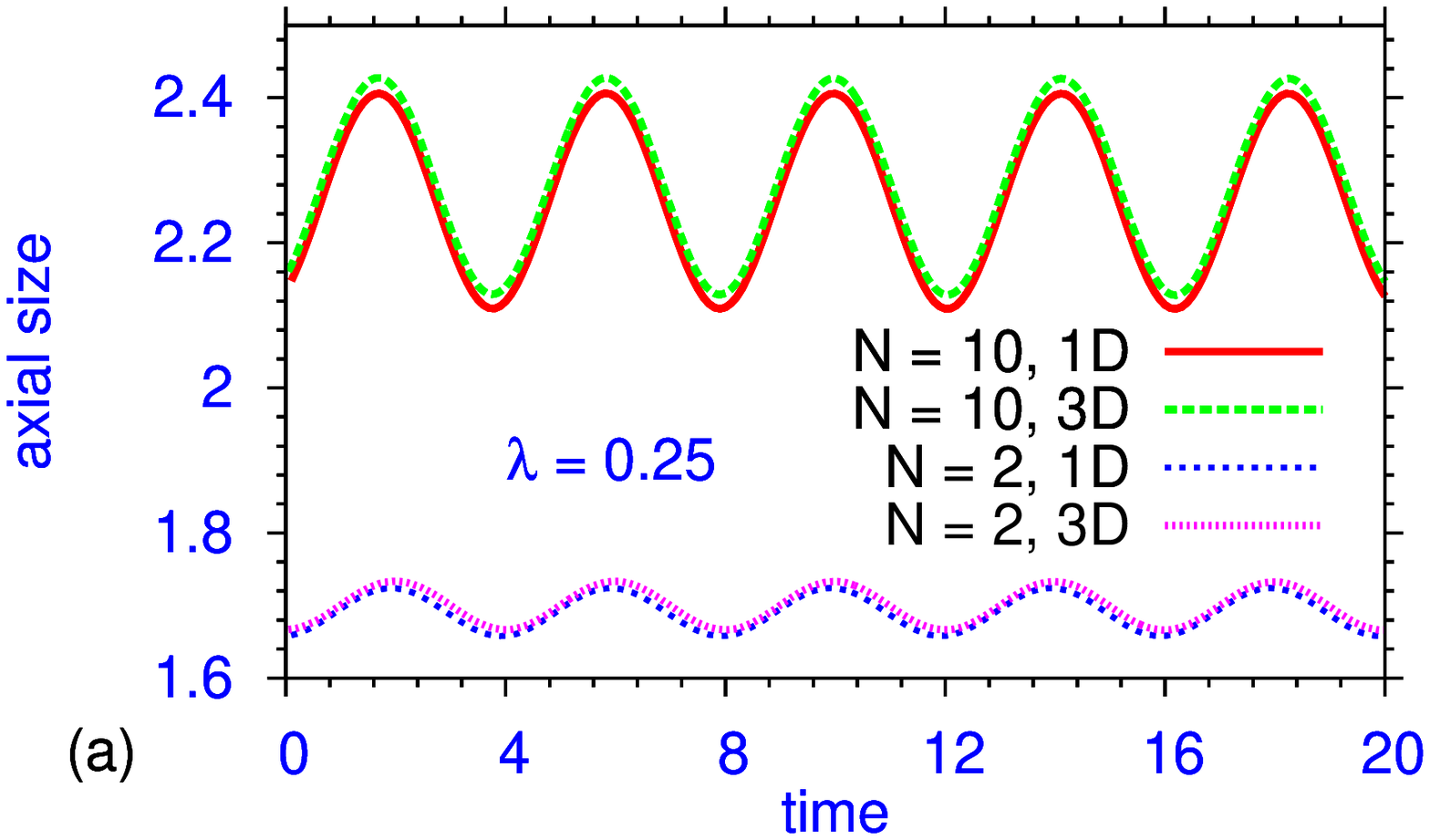}
\includegraphics[width=.49\linewidth]{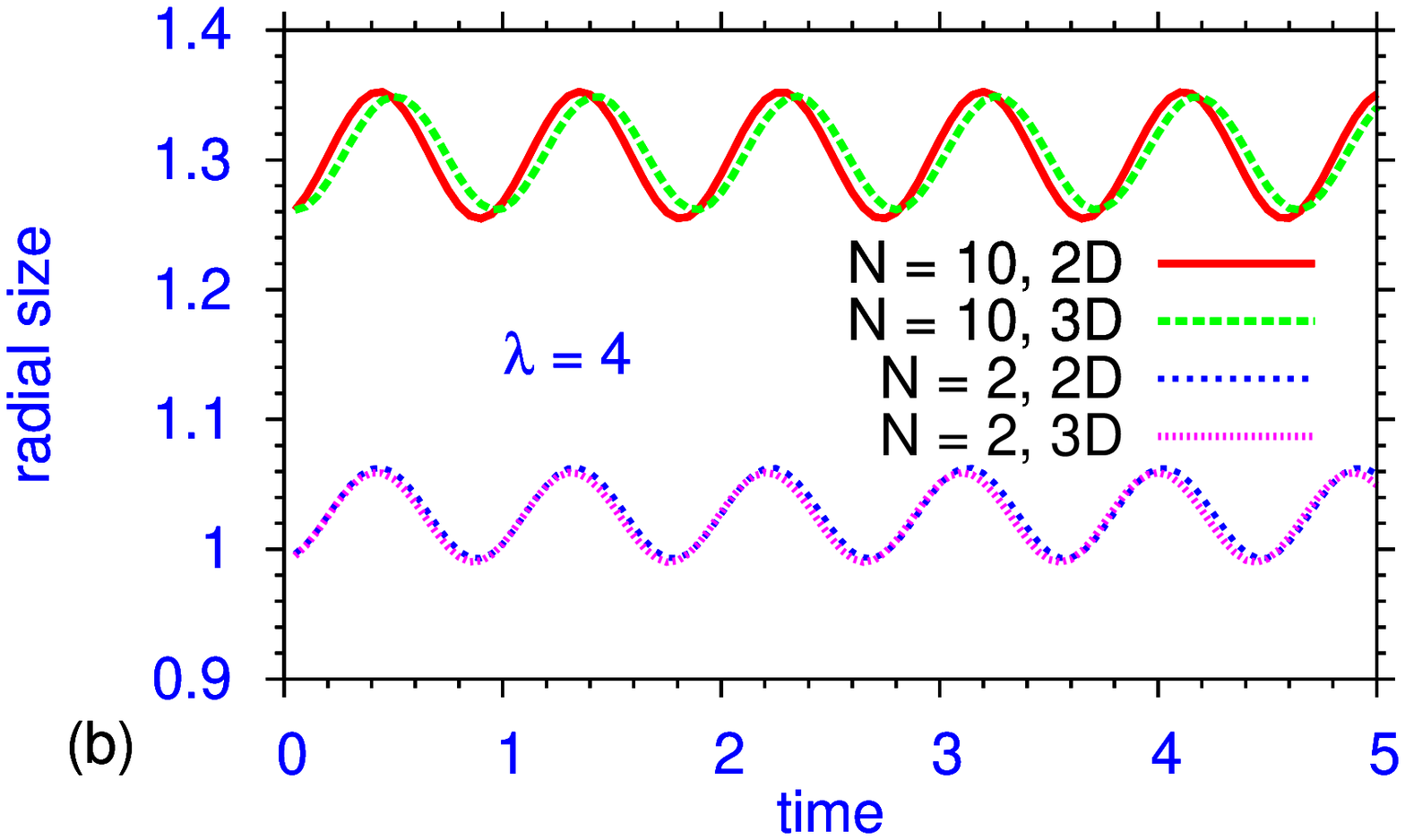}
\end{center}
\caption{(Color online) (a)  The rms axial size vs. time
(both in dimensionless oscillator units) during 
oscillation of a 
cigar-shaped
Fermi 
superfluid at unitarity ($\xi = 0.44$)
for $\lambda=0.25,$ started by reducing the axial trap 
suddenly by a factor of 0.9, as calculated by the full 3D equation 
(\ref{xe2}) and the reduced 1D equations (\ref{xe7}) and (\ref{xe12}) 
for 
$\alpha =0.98/\sqrt 2$.  (b) 
 The rms radial size vs. time 
(both in dimensionless oscillator units) during oscillation 
 of a disk-shaped
Fermi
superfluid at unitarity 
for $\lambda=4,$ started by reducing the radial trap
suddenly by a factor of 0.9, as calculated by the full 3D equation
(\ref{xe2}) and the reduced 2D equations (\ref{xe15}) and 
(\ref{xe22})
for
$\eta =0.92/\sqrt 2$.
}
\label{fig4}
\end{figure}

\subsection{Dynamics: Breathing oscillation}

Now we subject the reduced 1D and 2D
DF equations to a more stringent test, e.g., how well these equations 
can reproduce non-stationary (non-equilibrium) dynamics of a cigar- and  
disk-shaped Fermi superfluid. 
First we consider a cigar-shaped Fermi superfluid with $\lambda = 0.25$,
which is set into breathing oscillation by reducing only the axial
potential suddenly by a factor of 0.9. The radial trap is left 
unchanged. The resultant oscillation is
studied using the full 3D DF equation (\ref{xe2}) as well as the reduced
1D DF equation (\ref{xe7}) using the chemical potential (\ref{xe12}).
The root mean square (rms) axial size as calculated from the 3D and 1D
equations are plotted in figure \ref{fig4} (a). 
Next we consider a disk-shaped Fermi superfluid with $\lambda = 4$,
which is set into breathing oscillation by reducing only the radial
potential suddenly by a factor of 0.9. The resultant oscillation is
studied using the full 3D DF equation (\ref{xe2}) as well as the reduced
2D DF equation (\ref{xe15}) using the 
chemical potential (\ref{xe22}).
The rms radial size as calculated from the 3D and 2D
equations are plotted in figure \ref{fig4} (b). 

\subsection{Dynamics: Free expansion}

\begin{figure}

\begin{center}
\includegraphics[width=.49\linewidth]{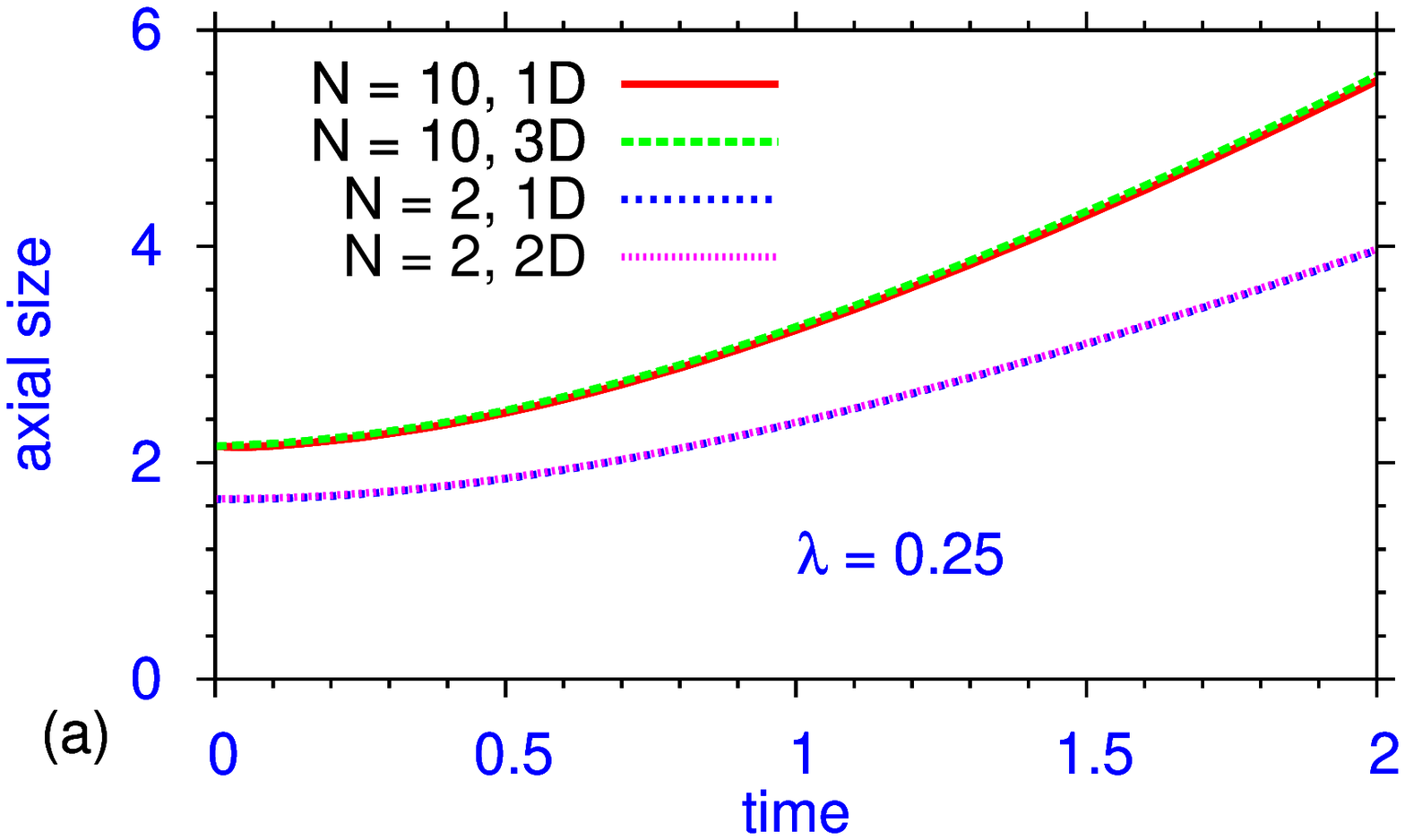}
\includegraphics[width=.49\linewidth]{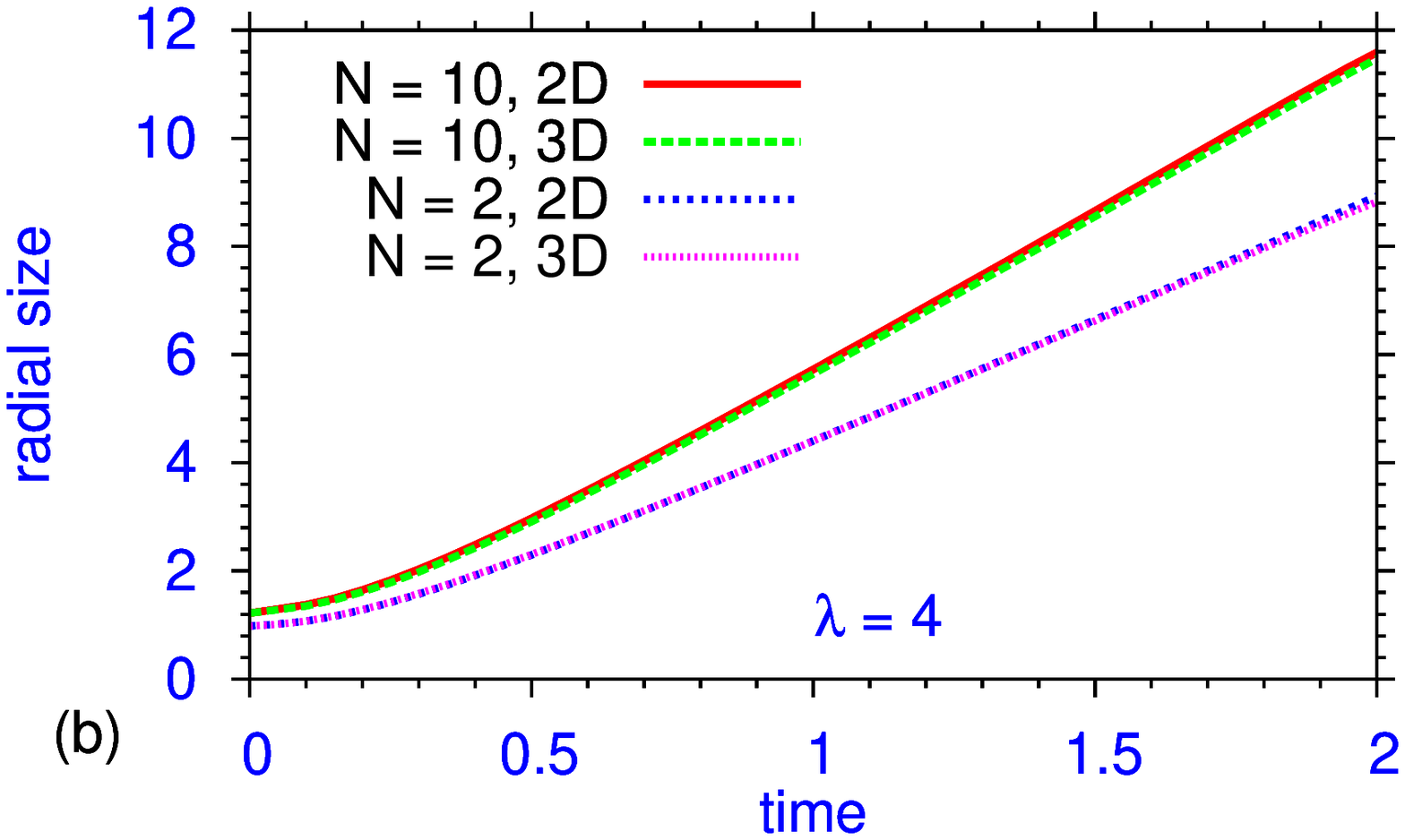}
\end{center}
\caption{(Color online) (a)  The rms axial size vs time 
(both in dimensionless oscillator units)
during axial  expansion  of a cigar-shaped
Fermi 
superfluid at unitarity ($\xi = 0.44$)
for $\lambda=0.25,$ started by removing  the axial trap 
suddenly, as calculated by the full 3D equation 
(\ref{xe2}) and the reduced 1D equations (\ref{xe7}) and (\ref{xe12}) 
for 
$\alpha =0.98/\sqrt 2$.  (b) 
 The rms radial size vs time
(both in dimensionless oscillator units)
during radial  
expansion of a disk-shaped
Fermi
superfluid at unitarity
for $\lambda=4,$ started by removing the radial trap
suddenly, as calculated by the full 3D equation
(\ref{xe2}) and the reduced 2D equations (\ref{xe15}) and 
(\ref{xe22})
for
$\eta =0.92/\sqrt 2$.
}
\label{fig5}
\end{figure}

Now we consider the problem of free expansion of a cigar- and 
disk-shaped Fermi superfluid, respectively, when the axial and radial 
traps are suddenly removed after the formation of the superfluid.
First we consider a cigar-shaped Fermi superfluid with $\lambda =0.25$ 
which is allowed to expand freely in the axial direction by setting the 
axial trap suddenly to zero in the 3D equation. The radial trap is left 
unchanged. The resultant expansion is studied using 
the full 3D DF equation (\ref{xe2}) as well as the reduced
1D DF equation (\ref{xe7}) using the chemical potential (\ref{xe12}).
The rms axial size as calculated from the 3D and 1D equations 
are plotted in figure \ref{fig5} (a). 
Next we consider a disk-shaped Fermi superfluid with $\lambda = 4$, 
which is allowed to expand freely in the radial direction by setting the 
radial trap suddenly to zero in the 3D equation. The axial trap is left 
unchanged. The resultant expansion is studied using
the full 3D DF equation (\ref{xe2}) as well as the reduced
2D DF equation (\ref{xe15}) using the chemical potential 
(\ref{xe22}). The rms radial size as calculated from the 3D and 2D 
equations
are plotted in figure \ref{fig5} (b).
The agreement between the dynamics as obtained from the full 3D
DF equation
and that from the reduced DF equations in Figs. \ref{fig4} and \ref{fig5}
is quite satisfactory.

\section{Conclusion}

In conclusion, we have suggested time-dependent mean-field reduced 
DF equations in 1D and 2D, respectively,
for a
cigar- 
and disk-shaped BEC and 
Fermi superfluid in the BCS and unitarity limits 
with simple analytic 
nonlinear terms. The numerical solution 
of these reduced equations reveals that they produce results for density 
of Fermi superfluids and BEC
in cigar- and disk-shaped traps in 
excellent 
agreement with the solution of the full 3D DF equation.  
We also studied non-stationary breathing oscillation  
of the cigar- and 
disk-shaped  Fermi superfluid  
initiated by a sudden change of axial and radial traps, respectively. Finally,  
we applied the reduced equations to the study of free expansion of 
a cigar- and disk-shaped Fermi superfluid initiated by a sudden removal 
of the axial and radial trap, respectively.   The reduced equations 
produced equally good results in both these studies when compared with the 
solution of the full 3D equations.

\ack

FAPESP, CAPES  and CNPq (Brazil) provided partial support.

\end{document}